\documentclass{osa-article}

\journal{oe}
\articletype{Research Article}
\begin{document}

\title{Fully refractive adaptive optics fluorescence microscope using an optofluidic wavefront modulator}

\author{Pouya Rajaeipour\authormark{1,*}, Alex Dorn\authormark{1}, Kaustubh Banerjee\authormark{1}, Hans Zappe\authormark{1}, and \c{C}a\u{g}lar Ataman\authormark{1}}

\address{\authormark{1}Gisela and Erwin Sick Laboratory for Micro-optics, Department of Microsystems Engineering, University of Freiburg, Germany}

\email{\authormark{*}pouya.rajaeipour@imtek.uni-freiburg.de} %% email address is required



\begin{abstract}
Adaptive optics (AO) is a powerful image correction technique with proven benefits for many life-science microscopy methods. However, the complexity of adding a reflective wavefront modulator and a wavefront sensor into already complicated microscope has made AO prohibitive for its widespread adaptation in microscopy systems. We present here the design and performance of a compact fluorescence microscope using a fully refractive optofluidic wavefront modulator yielding imaging performance on par with that of conventional deformable mirrors, both in correction fidelity and articulation. We combine this device with a modal sensorless wavefront estimation algorithm that uses spatial frequency content of acquired images as a quality metric and thereby demonstrate a completely in-line adaptive optics microscope which can perform aberration correction up to 4$^{th}$ radial order of Zernike modes. This entirely new concept for adaptive optics microscopy may prove to extend the performance limits and widespread applicability of AO in life science imaging.
\end{abstract}

%%%%%%%%%%%%%%%%%%%%%%%%%%  body  %%%%%%%%%%%%%%%%%%%%%%%%%%

\section{Introduction}
\label{sec:intro}
Adaptive Optics (AO) microscopy aims to enhance imaging quality by removing the aberrations resulting from refractive index inhomogeneities in a sample and component imperfections in a microscope~\cite{tyson-ao,liu2018observing}. However, large bulky optics and excessive optical complexity represent two major limitations of conventional solutions for realizing AO in microscopy~\cite{dubose2018handheld}.

In particular, conventional AO systems require a wavefront sensor for direct aberration measurement and a reflective wavefront modulator for phase error compensation~\cite{azucena2011adaptive,wang2014rapid,wang2015direct}. But in many microscopy techniques, due to physical and/or practical reasons, employing a wavefront sensor is not feasible~\cite{booth2014adaptive,ji2017adaptive,kubby-bio-ao}. Numerous effective solutions for indirect estimation of the aberrations without the need for a wavefront sensor have been proposed~\cite{booth2006wave,ji2010adaptive,linhai2011wavefront,vzurauskas2019isosense}, yet the cost and system complexity associated with integration of a reflective wavefront modulator into already complicated microscopes still remain significant challenges.

A few attempts at realizing AO microscopes utilizing liquid crystal wavefront modulators~\cite{bagwell2006liquid,tanabe2016transmissive} or piezoelectric multi-actuator lenses~\cite{bonora2015wavefront,two-photon-scanning,philipp2019diffraction} have been undertaken, but both of these solutions suffer from severe limitations: refractive LC modulators, despite their versatility due to the dense pixelation, have low transmission efficiency; polarization dependency; and higher noise due to diffraction artifacts. Piezoelectric lenses require sophisticated control strategies due to hysteresis effects and have limited actuator count which restricts the possible order of aberration correction. Their inability to correct for spherical and/or higher-order radially symmetric aberrations is a major limitation, since these aberrations are a major impediment for achieving high quality imaging in life-science microscopy. 

We have recently demonstrated an all-in-line AO microscope featuring a novel optofluidic refractive wavefront modulator which has a performance comparable to deformable mirrors in terms of correction fidelity and articulation~\cite{dorn2019compact}. Here we show for the first time in details how this technology can be used for compact wide-field microscopy with fully in-line pupil-plane AO correction.
In the context of a full microscopy system, we prove that the refractive wavefront modulator~\cite{OFAO-AO}, combined with an open-loop control system~\cite{OFAO-control-ao} and a sensorless wavefront estimation algorithm, are a clear step forward towards an almost plug-and-play solution for AO microscopy. A modal decomposition-based sensorless wavefront estimation algorithm, which uses spatial frequency content of the images as a quality metric, provides the wavefront aberration information. We show that this microscope can perform aberration correction up to 4$^{th}$ radial order of Zernike modes, and offer substantial increase in image quality in the presence of sample-induced aberrations. The limits of imaging and aberration correction performance are explored using fluorescent micro-beads, a 1951 USAF target sample and cheek cells imaged behind a 3D nano-printed phase plate.
%While our focus in this manuscript is AO microscopy, documenting the performance of this new type of modulator would also be of considerable interest for researchers interested in beam shaping in general.

%Using an optofluidic refractive wavefront modulator with performance comparable to deformable mirrors in terms of correction fidelity and articulation combined with an open-loop control for its optimal actuation~\cite{OFAO-AO,OFAO-control-ao}, we show here how this technology can be used for compact wide-field microscopy with fully in-line pupil-plane AO correction. A modal decomposition-based sensorless wavefront estimation algorithm that uses spatial frequency content of the images as a quality metric provides the wavefront aberration information.
%We show that this microscope can perform aberration correction up to 4$^{th}$ radial order of Zernike modes, and offer substantial increase in image quality in the presence of sample-induced aberrations. The limits of imaging and aberration correction performance are explored using fluorescent micro-beads and cheek cells imaged behind a 3D nano-printed phase plate.

\section{In-line pupil-plane AO for microscopy}
\label{sec:briefing}

\begin{figure}[t!]
 \centering
 {\includegraphics[width=0.6\columnwidth]{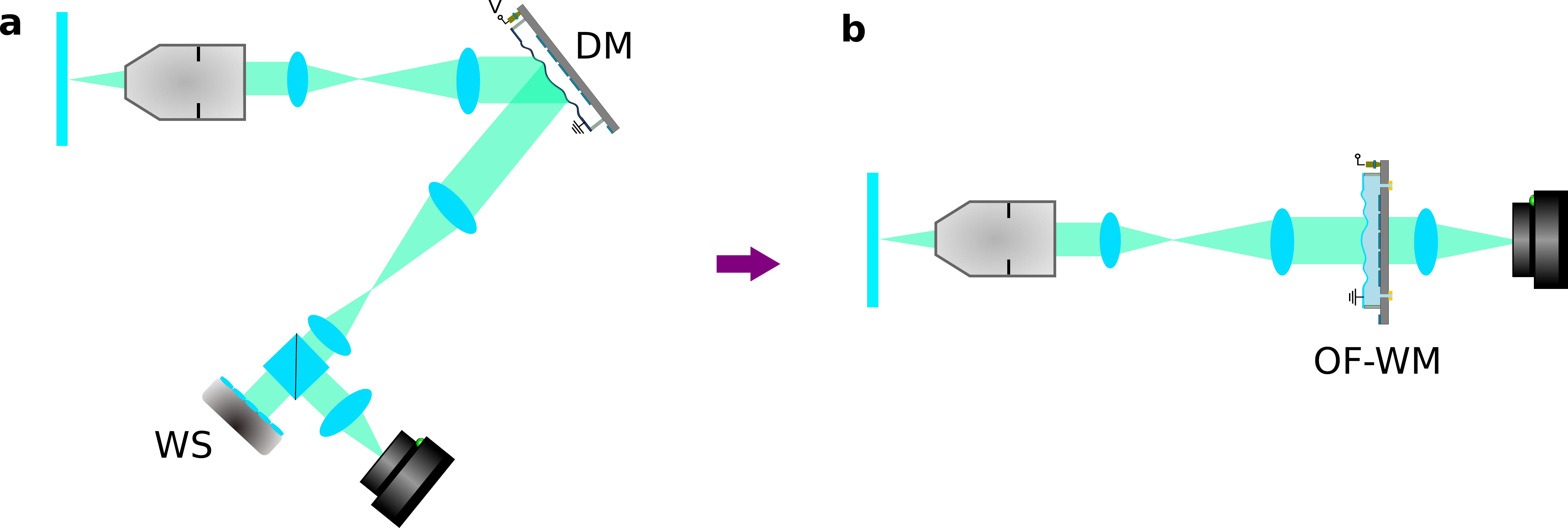}}
 \caption{Evolution of the conventional AO microscopy setups from: (\textbf{a}) requiring at least 2 telescopes and folded paths due to having a Deformable Mirror (DM) and a Wavefront Sensor (WS) to (\textbf{b}) the proposed configuration in this study which employs an Optofluidic refractive Wavefront Modulator (OF-WM) and requires only one telescope.}
 \label{fig:ao-architectures}
\end{figure}

The most common AO configuration is pupil-plane AO, where the wavefront correcting element is located at a plane conjugate to the microscope objective's pupil plane. Figure \ref{fig:ao-architectures}a depicts the simplified architecture of a conventional pupil-plane AO microscope. This system requires a reflective wavefront modulator (commonly a Deformable Mirror (DM)) and a Wavefront Sensor (WS) to directly measure the aberrations. To integrate such an AO module into a microscope at least two telescopes are necessary to relay the microscope pupil plane to the locations of the DM and WS. Furthermore, the optical path folds twice to accommodate the DM and the sensor. This configuration can rapidly become much more complex for scanned microscopy, due to additional telescopes and folded paths to integrate the scanning mirrors.

Figure \ref{fig:ao-architectures}b shows the all-in-line, purely refractive AO microscope architecture demonstrated here. Such an arrangement is only possible with a refractive wavefront modulator, for which only a single telescope is sufficient to scale the objective pupil diameter to that of the modulator. To avoid the need for a WS and an additional beam splitter, the wavefront modulator has to be operated in open-loop which requires hysteresis-free performance. This considerably simpler optical arrangement removes many of the complexities currently hindering the application of AO microscopy.

\subsection{AO fluorescence microscope design}
\label{sec:microscope}
\begin{figure}[t!]
 \centering
 {\includegraphics[width=0.70\columnwidth]{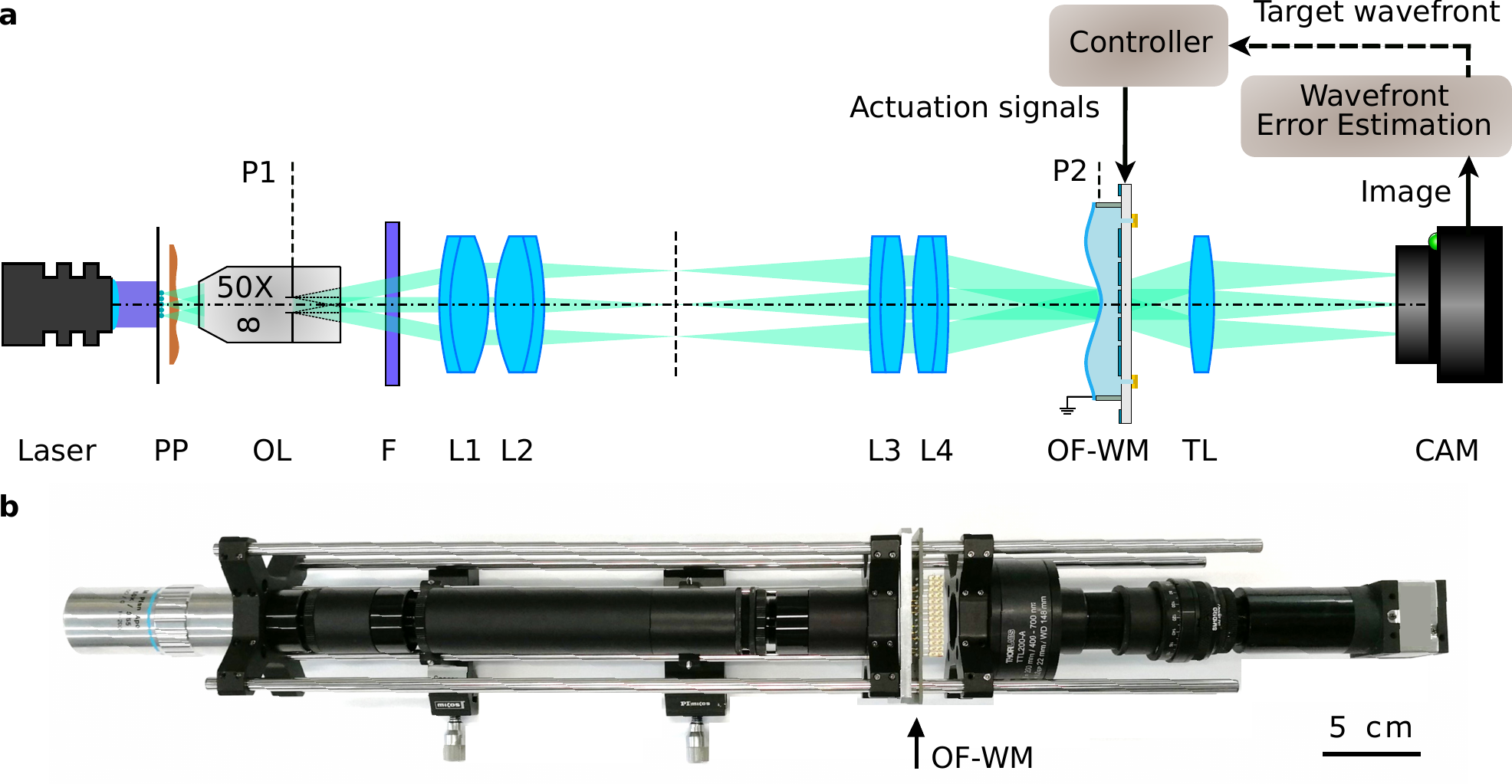}}
 \caption{Compact and all-in-line fluorescence microscope with sensorless AO aberration correction. (\textbf{a}) Schematics of the system design and (\textbf{b}) the realized AO microscope. PP: phase plate, OL: objective lens, F: filter, L1-L4: telescope lenses, OF-WM: optofluidic wavefront modulator, TL: tube lens.}
 \label{fig:ao-microscope}
\end{figure}

The layout of the fluorescence AO microscope using an in-line refractive wavefront modulator is shown in figure \ref{fig:ao-microscope}a. It features an infinity-corrected long working distance objective (Mitutoyo, 50X, NA=0.55, working distance=\SI{13}{\mm}) and the wavefront modulator is inserted in a plane conjugate to the pupil plane of the objective. The exit pupil of this objective has a diameter of \SI{4.4}{\mm} and lies inside the barrel of the objective (in this case \SI{14.9}{\mm} from its exit). Therefore, a Keplerian telescope is used to relay and scale the objective pupil onto the modulator. The clear aperture of the wavefront modulator intended for wavefront correction is \SI{9}{\mm}. A tube lens with a focal length of \SI{200}{\mm} (Thorlabs, TTL200-A) is then used to image the collimated light on the sensor of a CMOS camera (iDS, UI-1240SE-NIR-GL). The camera has a $1/1.8$ inch sensor with $1280\times1024$ pixels each with a size of \SI{5.3}{\um}.

The following design goals were considered in the conceptualization of the microscope:
\begin{enumerate}
\item achieving diffraction-limited performance over the entire Field of View (FoV) (restricted by the area of the CMOS sensor) at the fluorescence emission wavelength of \SI{485}{\nm};
\item imaging and 2X magnification of the objective pupil plane to cover the entire clear aperture of the wavefront modulator; and
\item minimizing the length of the entire system to achieve a compact and all-in-line microscope.
\end{enumerate}

The main factor limiting the microscope's optical performance is the telescope, since lenses with short focal lengths and large beam diameters lead to significant coma and astigmatism. A popular solution to maintain the compactness of the system and also avoiding lenses with short focal lengths is the use of a symmetric achromatic doublet configuration, also known as Pl{\"o}ssl eypiece~\cite{plossl-config}. Symmetrically combining 2 achromatic doublets with a small air gap between them results in an Effective Focal Length (EFL) that is nearly half of the EFL of each individual achromatic doublet. For the current microscope, we chose two pairs of commercially available achromatic doublets with focal lengths of \SI{100}{\mm} and \SI{200}{\mm} (Thorlabs, AC254-100-A and AC254-200-A) which yields a total telescope length of \SI{300}{\mm}. The relay lenses are assembled inside a 1 inch lens tube. Figure~\ref{fig:ao-microscope}b shows the completely assembled adaptive optics microscope with a total length of \SI{590}{\mm}.

\subsection{Wavefront modulator and open-loop control}
\label{sec:PM}
The adaptive element used in the microscope is an optofluidic wavefront modulator~\cite{OFAO-AO,OFAO-OMN2018}. Similar to DMs, the modulator features a 2D array of electrostatic actuators distributed across the optical aperture, providing accurate replication of high-order aberrations. Unlike DMs, however, it is a refractive component that operates in transmission with high efficiency.
It comprises a glass substrate bearing a 2D electrode array, and a ring spacer and a thin polymer membrane forming a fluidic chamber. The chamber is filled with an incompressible optical liquid. The membrane deforms with voltage applied to the electrodes, locally changing the amount of liquid in the optical path. By controlling these local deflections, the shape of a wavefront refracted through the device can be modulated. Due to its hysteresis-free behavior, the modulator can be driven in open-loop using an optimization-based control strategy~\cite{OFAO-control-ao}. It has been shown that this wavefront modulator is capable of reproducing up to 5$^{th}$ order Zernike polynomials with high fidelity. Zernike modes are described here using the OSA/ANSI standard numbering scheme~\cite{thibos2002standards}, without normalization of the amplitudes.

\subsection{Sensorless wavefront estimation}
\label{sec:MWE}
In the absence of a discrete wavefront sensor, an algorithmic approach that directly uses recorded images becomes necessary for determination of the wavefront. We employed here a \textit{modal decomposition} algorithm adapted from Booth et al.~\cite{booth-lsf,vzurauskas2019isosense} in which the wavefront aberrations are represented as a linear combination of an orthonormal set of basis functions (modes), and the weighting coefficient of each mode is estimated independently. The wavefront estimation is achieved by maximizing a chosen image quality metric by cancelling the effects of each individual aberration mode, Zernike modes in our case. Our image quality metric is related to image sharpness~\cite{booth-lsf}. For each image, this metric is calculated by the integration of the normalized Power Spectral Density (PSD) between spatial frequencies 1\% to $\sim$15\% of the cut-off frequency, which is defined as the spatial frequency where the diffraction-limited MTF diminishes. This range is chosen such that the very low and high frequencies are ignored to compensate for the image intensity changes and noise, respectively.

\begin{figure}[t!]
 \centering
 {\includegraphics[width=0.95\columnwidth]{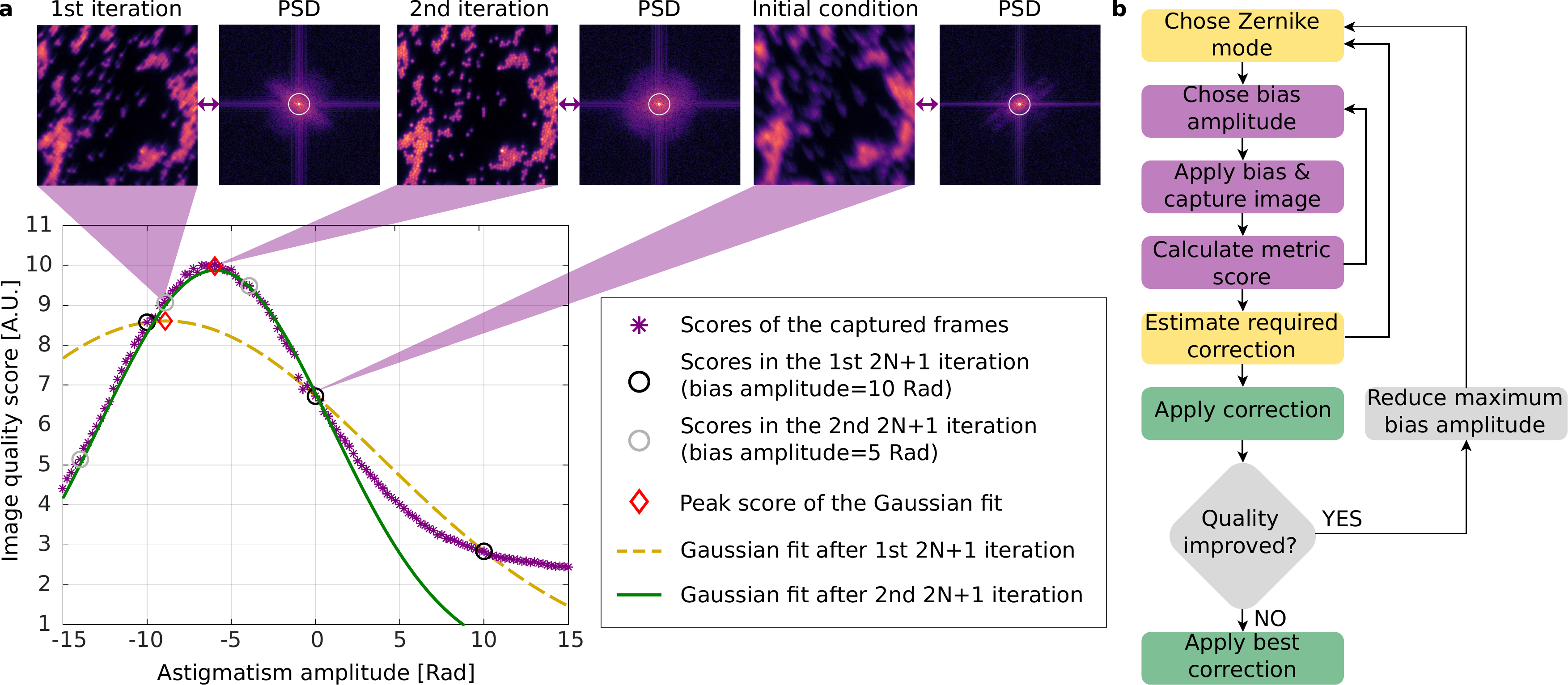}}
 \caption{(\textbf{a}) Sample implementation of the $2N+1$ algorithm for estimating an aberration mode with 2 iterations (in this case $N=1$). Data points are calculated by applying an additional bias aberration in the form of the analysed mode using the wavefront modulator and scoring the captured images. Asterisks depict the scores of the captured images by sweeping the amplitude of the applied oblique astigmatism. The quality metric spatial frequency range is shown on the PSD plots as an annular ring. (\textbf{b}) Summarized flowchart of the developed modal decomposition-based aberration estimation algorithm.}
 \label{fig:MWE}
\end{figure}

The algorithm used for aberration estimation is outlined in figure \ref{fig:MWE} and summarized as a flowchart in figure \ref{fig:MWE}b. To analyze the performance of this method, the wavefront modulator itself is used to introduce known aberrations at the pupil plane. The basic assumption underlying this algorithm is that a contrast-based image quality metric is a convex function of the amplitude of an aberration mode for an orthonormal set such as the Zernike modes, with a single maximum corresponding to the required correction for a specific aberration mode~\cite{booth2002adaptive}. This behavior is experimentally demonstrated in figure \ref{fig:MWE}a, which plots the image quality metric for a sample image (\SI{1}{\micro\meter} in diameter fluorescent beads) as a function of astigmatism amplitude. To obtain this behavior, the wavefront modulator is used to apply biases of oblique astigmatism ($Z_2^{-2}$) with amplitudes ranging from \SI{-15}{\radian} to \SI{15}{\radian} in steps of \SI{0.25}{\radian}.

In a practical application, however, obtaining such curves for all relevant aberration modes is impractical due to the large number of images required, so that the quality metric is measured at only three points (an initial condition and two bias aberrations on either side of it). The maximum of a convex curve is then fitted to these three measurements yielding an estimated amplitude for that aberration mode. Figure \ref{fig:MWE}a also plots such a Gaussian fit to three points at 0,+10 and \SI{-10}{\radian}. Since the actual function is not strictly Gaussian, the maximum of the fitted curve deviates from the actual value, leading to a poor estimation.

To overcome this problem, we employ multiple fitting iterations with progressively smaller bias amplitudes. As depicted in figure \ref{fig:MWE}a, the correction estimated from the first iteration is taken as the initial condition for the second iteration of the algorithm, for which we apply smaller biases ($\pm$\SI{5}{\radian}). The solid line in figure \ref{fig:MWE}a shows the Gaussian fit obtained from the second iteration, whose maximum this time closely approximates the required correction. We perform a minimum of two and maximum of five iterations of the algorithm with biases of $\pm10$ rad to $\pm1$ rad in the final iteration. Each iteration of the algorithm requires $2N+1$ measurements, where $N$ is the number of aberration modes considered for correction.

Since the procedure above needs to be repeated for all the relevant modes, the resulting measurement time can be long. However, this iterative approach has two major advantages that render it a viable option:
\begin{enumerate}
\item The Gaussian fit is more accurate when the applied biases are smaller and closer to the maxima.
\item The Zernike modes are not necessarily orthonormal with respect to the chosen quality metric specially for large amplitude aberrations~\cite{sensorless-accuracy,kazasidis2018algorithm}. This existing crosstalk between the influence of different aberration modes on the quality metric function leads to inaccuracies for estimating the real optimum wavefront correction. Therefore, executing multiple iterations of the algorithm with large initial biases, reduces the overall wavefront error in the primary iterations, and thus, weakens the present crosstalk in the latter iterations.
\end{enumerate}

\section{AO microscope optical performance}
\label{sec:exp-eval}

\subsection{Imaging characteristics}
\label{sec:img-prop}

As test samples for characterization, we used \SI{1}{\micro\meter} in diameter fluorescent beads dispensed on a microscope slide and a 1951 USAF target. A laser diode (Roithner LaserTechnik, RLDE405-12-6) with a center wavelength of \SI{405}{\nm} was used to excite fluorescent beads (Polysciences, Fluoresbrite YG Microspheres \SI{1.00}{\um}) with flood illumination from the backside. The fluorescence emission at the wavelength of \SI{485}{\nm} was collected, while the excitation and environment light is blocked using an emission filter (Thorlabs, MF479-40).

The magnification of the developed AO microscope was determined by imaging the 1951 USAF target sample. The emitted light of the fluorescent beads was used to illuminate the target sample at the design wavelength of the microscope. Considering the known width of the USAF target line elements and by calculating Full Width at Half Maximum (FWHM), the overall magnification of the microscope is calculated to be 25.0X in horizontal direction and 25.7X in the vertical direction.

\begin{figure}[t!]
 \centering
 {\includegraphics[width=0.86\columnwidth]{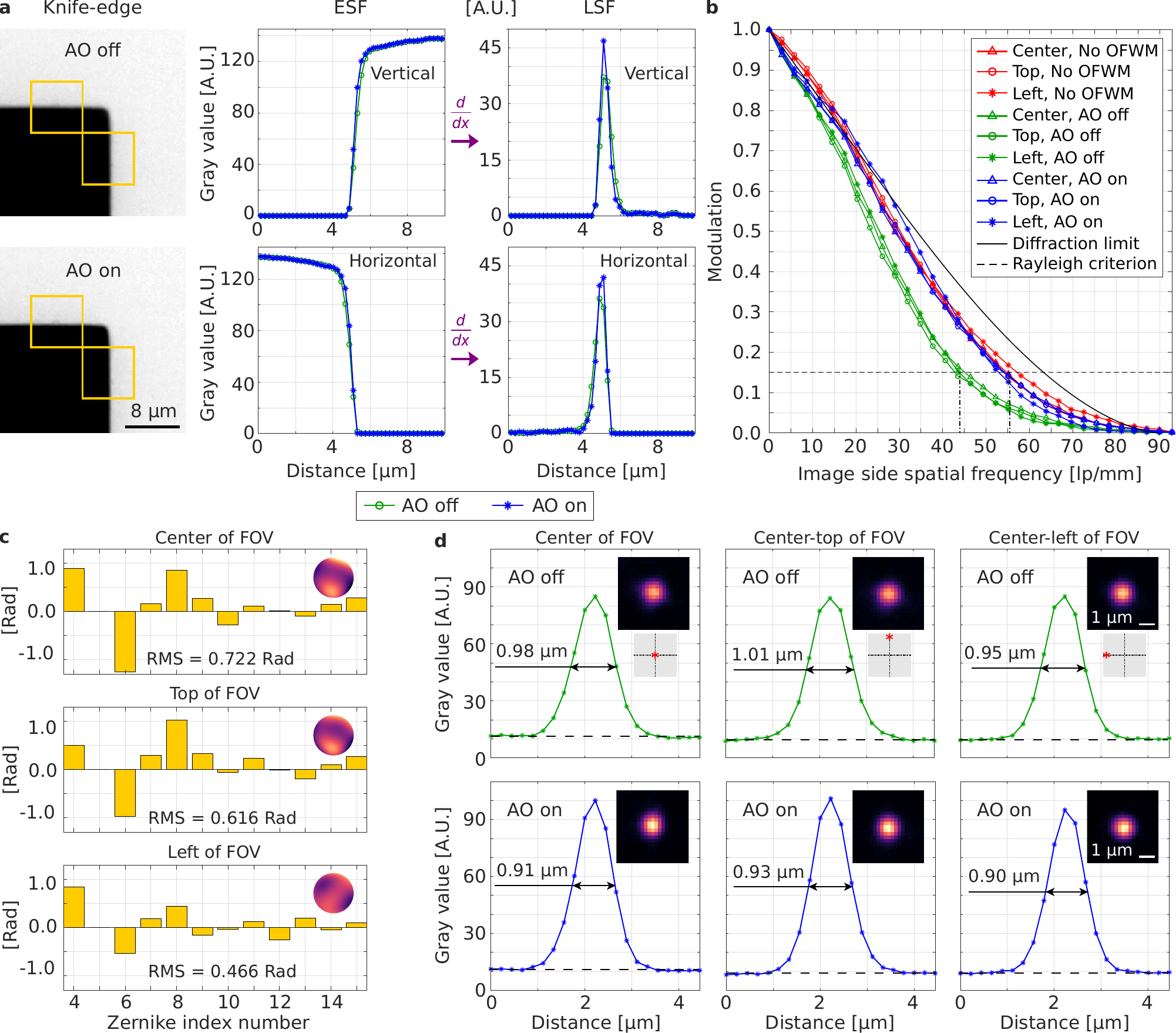}}
 \caption{(\textbf{a}) Knife-edge method for MTF measurement of the microscope thorough calculation of Edge Spread Function (ESF). (\textbf{b}) Image-space MTF of the developed AO microscope at different positions of the FoV before installing the optofluidic wavefront modulator (red), after installing it while it is all off (green) and after performing AO correction by sensorless estimation of the initial system aberrations (blue). (\textbf{c}) Zernike decomposition of the estimated initial aberrations at 3 different positions of the FoV. (\textbf{d}) Image of a single fluorescent bead with a nominal diameter of \SI{1}{\um} at different positions of the FoV before and after correcting for the estimated initial wavefront error of the fluorescence microscope. The line plots are calculated by averaging 4 equally-spaced cross-sections of the image of beads and transforming the distance to object space.}
 \label{fig:mtf-bead}
\end{figure}

The image-side Modulation Transfer Function (MTF)~\cite{mtf-measurement} is measured before and after installing the optofluidic wavefront modulator using the knife-edge method summarized in Figure \ref{fig:mtf-bead}a. Images of two perpendicular sharp edges aligned to respective frame borders are recorded (squares in figure \ref{fig:mtf-bead}a) at the desired location within the FoV. Several cross-sections over the edge are first averaged to obtain a noise-reduced Edge Spread Function (ESF). The derivative of the ESF provides the Line Spread Function (LSF), and the absolute value of Fourier transform of the LSF yields the MTF. The calculated MTFs in horizontal and vertical directions are averaged to obtain the overall MTF curve.

Summary of the MTF meausurement results are presented in figure \ref{fig:mtf-bead}b. The black curve depicts the theoretical diffraction-limit after compensating for the sensor point spread function (PSF)~\cite{mtf-measurement}, while the dashed-line corresponds to the Rayleigh limit for the minimum intensity modulation for resolving a pair of black and white lines. Red curves show the microscope MTF before installing the optofluidic wavefront modulator. Considering the magnification of the microscope, an average resolution of \SI{708}{\nm} in the object space over the entire FoV is achieved. The slight deviation from the diffraction limit at \SI{625}{\nm} is attributed to the misalignments in the telescope optics. After installing the wavefront modulator in its all off state, the MTF curves degrade (green curves) due to inherent deformations of the modulator membrane (detailed characteristics of the initial flatness of the employed wavefront modulator can be found in~\cite{OFAO-AO,OFAO-PW-pushpull}). By sensorless estimation and correction of the inherent modulator surface, the MTF of the microscope is recovered (blue curves). For this the image quality metric of the wavefront estimation algorithm is adapted for scoring the measured MTFs.

The bar-plots in figure \ref{fig:mtf-bead}c depict the estimated wavefront correction for respective portions of the FoV. The estimated aberrations, with an average of \SI{0.601}{\radian} RMS, are highly similar, which is expected since they are mostly induced by the inherent deformations of the wavefront modulator placed at the pupil plane. We also examined the performance of the developed AO module and the sensorless aberration estimation algorithm for correcting the initial system aberrations by imaging a single fluorescent bead with a nominal diameter of \SI{1}{\um} at different positions in the FoV. The total area of the FoV is $268\times 214$ $\mu m^2$. Figure \ref{fig:mtf-bead}d shows the image of a fluorescent bead, placed at the center of the FoV, before and after AO correction, and the line plots show the average gray value of the image of the bead at four different cross-sections. The x-axis is transformed to the object-space. The second and third columns of images in Figure \ref{fig:mtf-bead}d show the same data when the fluorescent bead is placed at the center-top and center-left of the FoV, respectively (shown by the red asterisks in the diagrams). The captured fluorescent intensity in the image of the beads after AO correction improves by 16.5\% on average.

\subsection{Wavefront error estimation characteristics}
\label{sec:exp-eval}
The aberration estimation algorithm was characterized by first correcting for known aberrations. For this, a set of randomly generated wavefront errors with increasing phase Root-Mean-Square Error (RMSE) including up to 4$^{th}$ radial order of Zernike modes were applied at the pupil plane of the microscope using the integrated wavefront modulator itself. The sensorless wavefront estimation algorithm was then utilized to estimate these aberrations.

\begin{figure}[t!]
 \centering
 {\includegraphics[width=0.95\columnwidth]{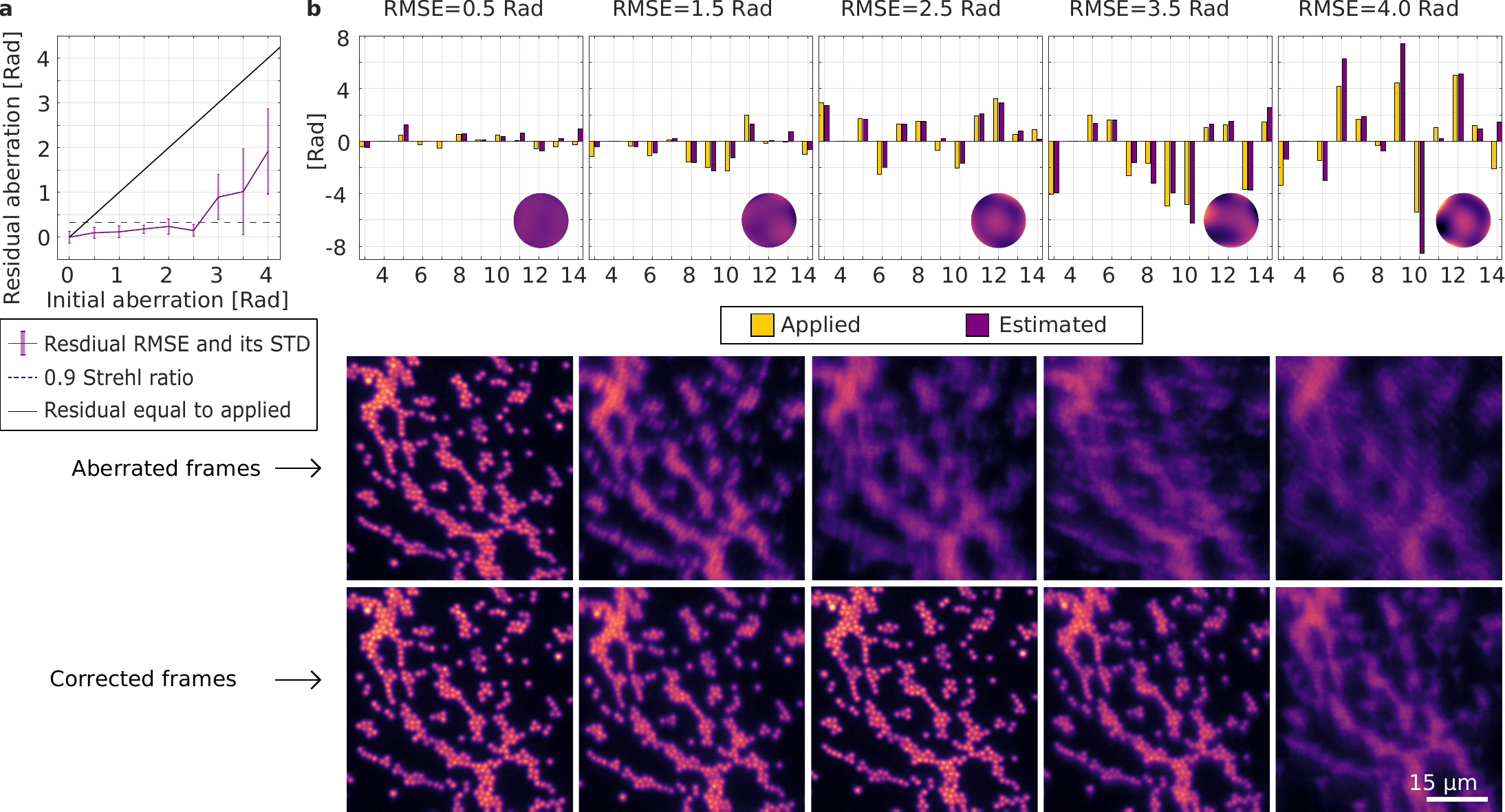}}
 \caption{Examples of correcting for known aberrations. (\textbf{a}) The line plot depicts the residual wavefront RMSE with respect to the applied aberrations after compensating for the system initial aberrations. (\textbf{b}) The bar plots show the applied and estimated aberrations. The first row of the figures show the frames that are aberrated by the wavefront errors shown in the bar plots. The bottom row frames show the corrected frames of each corresponding top aberrated frame.}
 \label{fig:MWE-charac-ex}
\end{figure}

The line plot in figure \ref{fig:MWE-charac-ex}a depicts the summary of this characterization by demonstrating the residual wavefront RMSE with respect to the applied aberrations after compensating for the system initial aberrations. Up to 5 iterations of the algorithm were used. The vertical bars at each magnitude of applied aberrations show the standard deviation of the residual wavefront errors. Each bar is calculated by correcting for 11 different aberrations. The dashed-line corresponds to the maximum permissible residual RMSE to conserve diffraction-limited imaging based on a Strehl ratio of 0.9. The straight solid line depicts a residual RMSE equal to the applied RMSE. Up to an RMSE of \SI{2.5}{\radian} the aberration estimation algorithm can retrieve the Strehl ratio to above 0.9. Furthermore, for larger aberrations up to \SI{4.0}{\radian}, the algorithm reduces the RMS aberration amplitude by more than 60\% on average. It should be noted that the available stroke of the wavefront modulator is limited in this experiment because part of it is used for inducing known aberrations. Therefore its actual correction capability is higher in actual specimen-induced aberration correction.

The bar plots in figure \ref{fig:MWE-charac-ex}b depict some examples of the performed corrections for known aberrations. The applied aberrations have an RMSE of \SI{0.5}{\radian} to \SI{4.0}{\radian} and the wavefront error for each magnitude of RMSE is chosen as the median of the set of aberrations used to generate the residual aberration characterization plot. The yellow and violet bars in the bar plots show the Zernike polynomial decomposition of the applied and estimated aberrations, respectively. The wavefront profiles on each bar plot correspond to the applied aberrations. The first row of pictures show the aberrated frames due to aberrations depicted in their corresponding top bar plots. The second row shows the resulting corrected frame after executing the aberration estimation algorithm.

\begin{figure}[t!]
 \centering
 {\includegraphics[width=0.9\columnwidth]{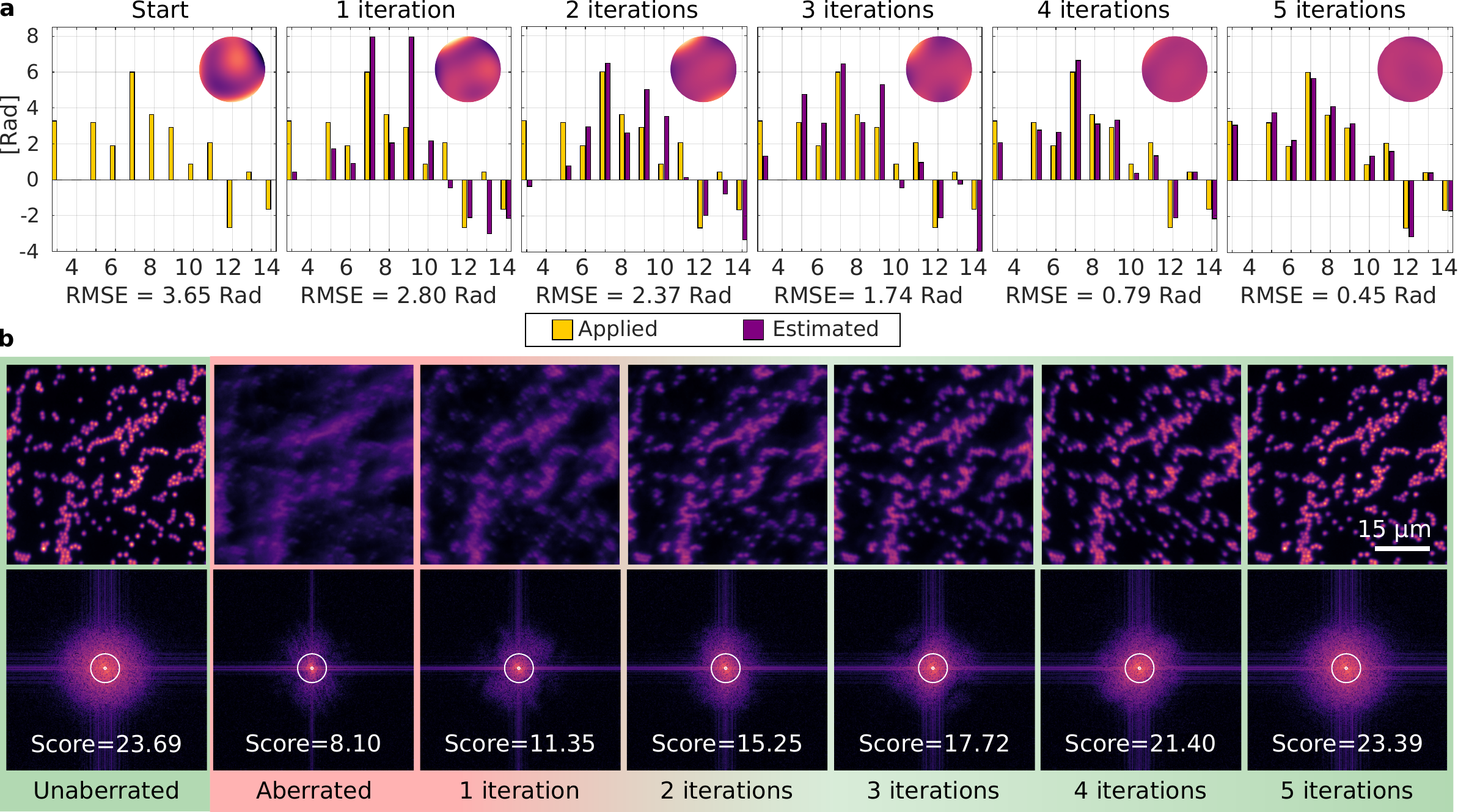}}
 \caption{Illustration of the progressive improvement in the image quality after each iteration of the aberration estimation algorithm. (\textbf{a}) The bar plots show the evolution of estimated wavefront towards the actual applied error. The residual RMSE after each iteration is depicted below the plots. (\textbf{b}) The captured frames and their PSD with corresponding score starting from the unaberrated condition towards the corrected frame after 5 iterations. The considered spatial frequency range for the image quality metric is depicted on the PSD plots in form of an annular ring.}
 \label{fig:itr-res}
\end{figure}

Figure \ref{fig:itr-res} shows the advantage of performing multiple iterations of the aberration estimation algorithm. The pictures in figure \ref{fig:itr-res}b depict the captured frames along with their PSD and their corresponding quality score. The spatial frequency range considered for the image quality metric is shown as an annular ring on top of the PSD plots. The yellow bars in the figure \ref{fig:itr-res}a depict the Zernike polynomial decomposition of an example wavefront error applied by the wavefront modulator at the pupil plane of the microscope. The violet bars show the estimated wavefront error after each iteration. The required number of iterations for each correction is based on: (i) image quality score improvement such that in case of a decrease in the quality score the iterations are stopped and (ii) the desired image quality. By executing 5 iterations of the algorithm the unaberrated image quality is retrieved. 

%%%%%%%%%%%%%%%%%%%%%%%%%%%%%%%%%%%%%%%%%%%%%%%%%%%%%%%%%%%%%%%%%%%%%%%%%%%%%%%%%%%%
\section{Aberration correction performance}
\label{sec:res}
The imaging performance of the developed AO-microscope is finally demonstrated by correcting for aberrated images of different objects with distinct spatial frequency content. To mimic sample-induced aberrations, a custom phase plate is manufactured via two-photon polymerization-based 3D printing. A transparent 1951 USAF target is positioned at the object plane of the microscope and this phase plate is directly placed on top. A wide-band white light source is used for illumination of the object from back-side while only a narrow-band of the illuminating light with center wavelength of 479 nm is passed through the microscope filter.
We have limited the sensorless aberration estimation and correction to up to 4$^{th}$ radial order of Zernike modes. The reason was twofold: firstly, for pupil-AO microscopy, any correction beyond the 4$^{th}$ order has diminishing returns, since most common aberrations can be represented within this range~\cite{schwertner2004characterizing}. On the other hand, the number of images necessary for each iteration of the sensorless estimation technique scales with $2N$, with more orders meaning longer acquisition time. We therefore chose $N=11$ as the optimum compromise.

\begin{figure}[t!]
	\centering
	{\includegraphics[width=0.74\columnwidth]{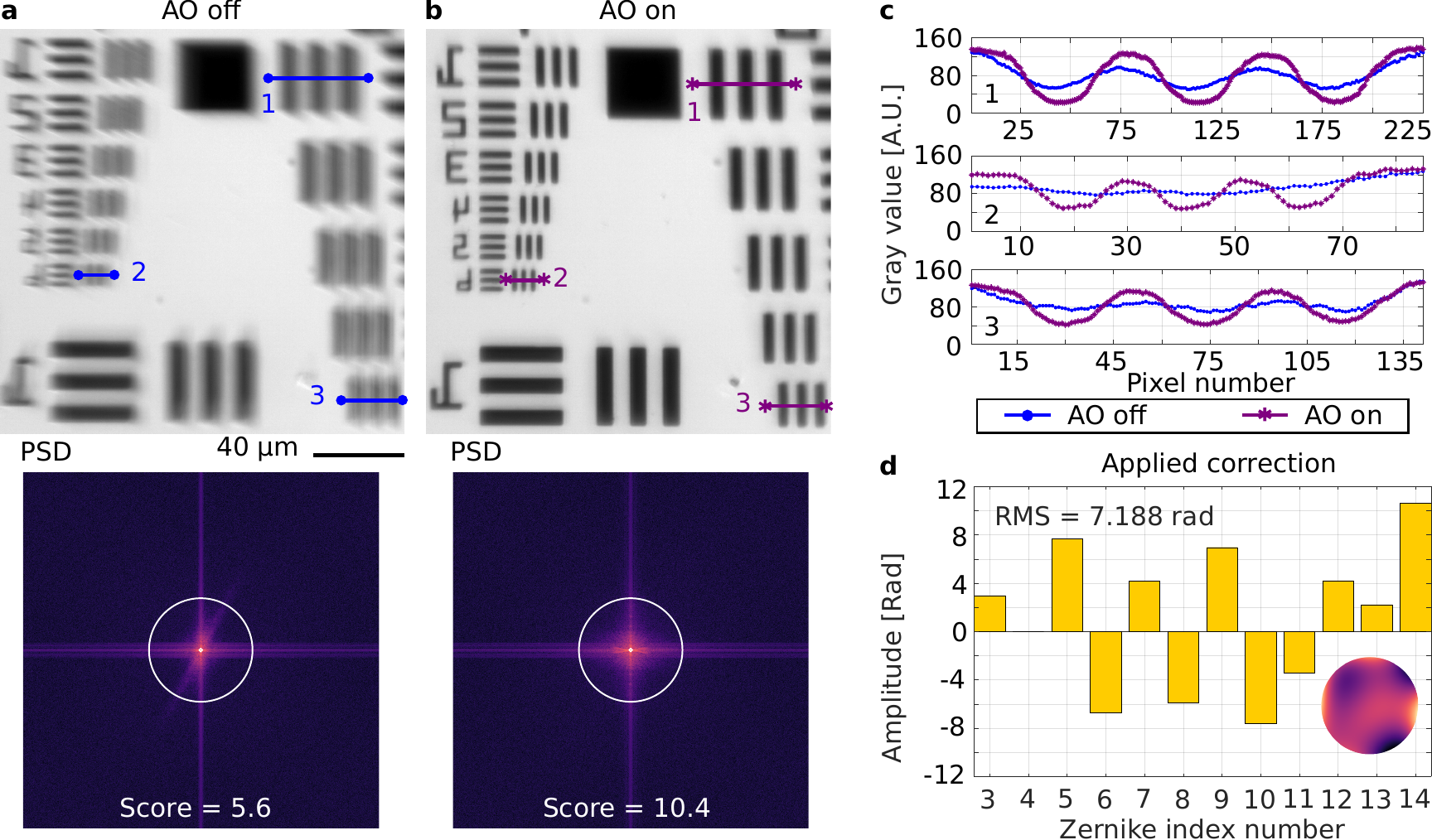}}
	\caption{(\textbf{a})~Aberrated image of a 1951 USAF target due to placing a custom phase plate on top of the target. (\textbf{b})~By estimating the aberrations and performing AO correction the image quality is retrieved and the PSD of the image includes higher spatial frequency content compared to the PSD of the aberrated image. Line elements at the right and top-left of the image correspond to groups 6 and 7 of the target. The annular rings on the PSD plots depict the spatial frequencies between 1\% to 35\% of the cut-off frequency, considered for the image quality metric. (\textbf{c})~Intensity profiles at 3 marked regions of the target before (circles) and after (asterisks) AO correction. (\textbf{d})~Zernike decomposition of the applied correction.}
	\label{fig:usaf-ao}
\end{figure}

\begin{figure}[t!]
	\centering
	{\includegraphics[width=0.92\columnwidth]{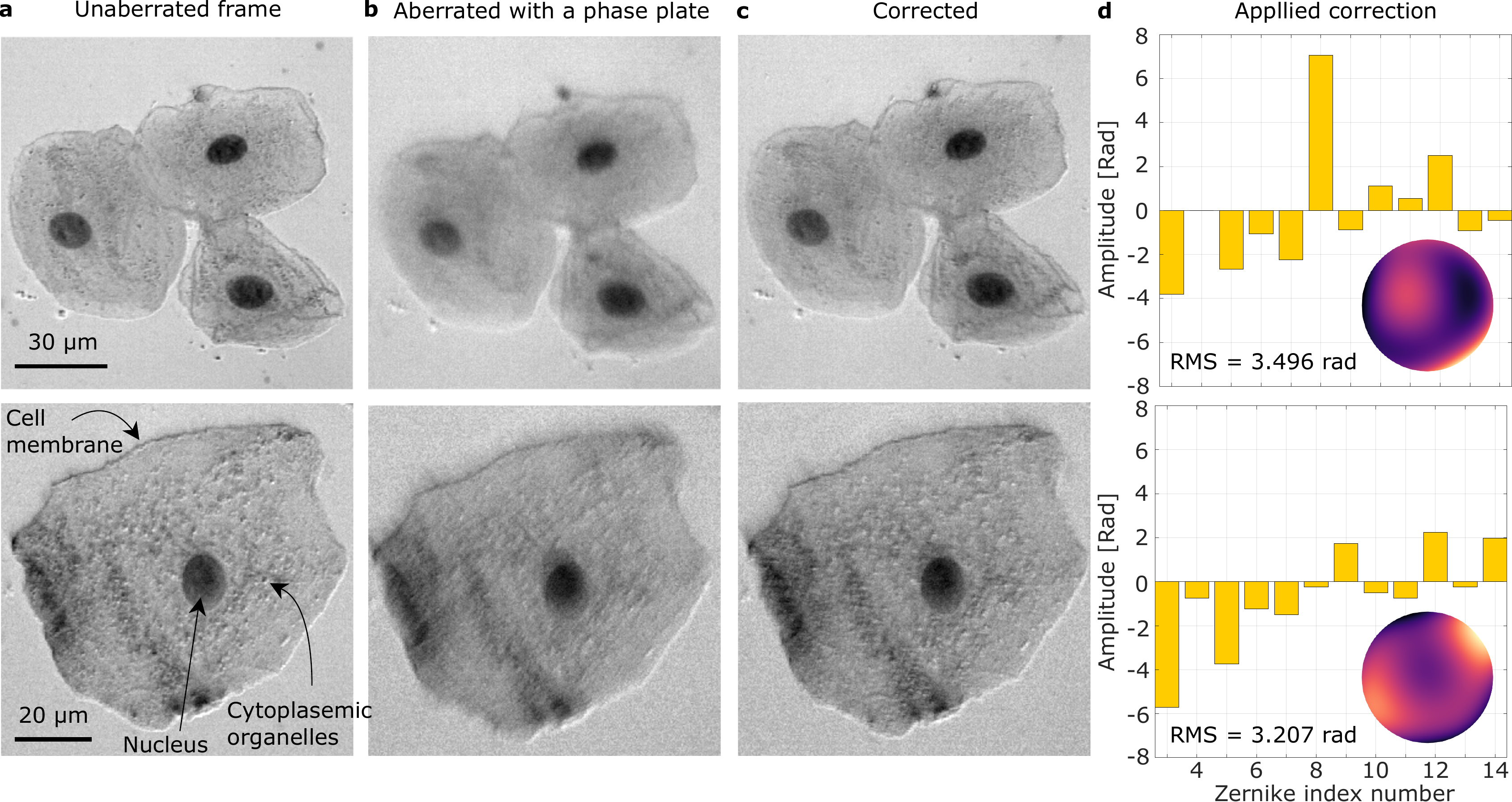}}
	\caption{Examples of imaging human cheek cells by developed AO-microscope. (\textbf{a})~Unaberrated images of the cells after correction for the initial wavefront modulator flatness error and potential spherical aberration due to the cover slip. (\textbf{b}) Aberrated images which are distorted by placing an aberrating phase plate on top of the cover slip. (\textbf{c}) The corrected images which retrieve the resolvability of the cheek cell details and (d) the Zernike decomposition of the applied corrections.}
	\label{fig:cheek-ao}
\end{figure}

Figure~\ref{fig:usaf-ao}a shows the aberrated image of the target and its PSD while the AO module is off. In this case, the upper limit of the spatial frequency range in the employed image quality metric is increased to 35\%, depicted as an annular ring on top of the PSD plots. The right and left groups of line elements in the image belong to the groups 6 and 7 of the USAF target, respectively. The modal wavefront estimation algorithm with 5 iterations is executed to estimate the aberrations using the complete area shown in the figure. Figure~\ref{fig:usaf-ao}b shows the corrected frame after applying the estimated correction demonstrating clear improvement in imaging quality over the entire FoV. The PSD of the corrected image compared to the aberrated one includes contents with higher spatial frequencies verifying that the fine details of the target are resolvable after correction. The intensity line plots in figure~\ref{fig:usaf-ao}c highlight the contrast and resolution of the image before and after applying AO correction over 3 different regions of the FoV. While elements of the group 7 of USAF target were initially not resolvable due to aberrations, they are clearly separable after AO correction. Figure~\ref{fig:usaf-ao}d outlines the Zernike decomposition of the estimated wavefront error with a RMSE of 7.188 rad.

To evaluate the performance of the AO microscope for use with biological samples, we imaged stained cheek cells. The cells were dispensed on a microscope slide, stained with black ink and covered by a thin cover slip. The upper limit of the spatial frequency range in the employed image quality metric was increased to 85\%. The same illumination as previously was used. Figure~\ref{fig:cheek-ao}a depicts the samples without the aberrating phase plate, for which the initial aberrations of the system are corrected by running the sensorless algorithm for the corresponding FoV. The initial aberrations are due to the inherent flatness error of the wavefront modulator and the potential spherical aberration that might arise due to the cover slip. A different area of the previously described phase plate was placed on top of the cover slip to induce optical aberrations, which mimics the aberrations for the case of deep tissue imaging. In the aberrated frames, shown in figure\ref{fig:cheek-ao}b, the details of the cheek cell structure including the cell membrane, the nucleus border and cytoplasmic organelles are lost. After performing AO correction, using sensorless estimation of the distorting aberrations, these details are clearly visible in the corrected frames shown in figure~\ref{fig:cheek-ao}c. The Zernike decomposition and wavefront profiles of the estimated aberrations are shown in figure~\ref{fig:cheek-ao}d.

\section{Conclusion}

We have presented a fully refractive wide-field fluorescence AO-microscope and documented its feasibility through imaging experiments on non-biological and biological samples under controlled aberrations. The novel architecture combining an optofluidic wavefront modulator alongside a modal decomposition based wavefront estimation algorithm eliminates the need for the folding of the optical path, drastically reducing the size, complexity and cost of typical AO microscope implementations. 

\section*{Acknowledgments}
This work is partially funded by Deutsche Forschung Gemeinschaft (DFG) (Project 62751: OFAO).

\section*{Disclosures}
The authors declare no conflicts of interest.

%%%%%%%%%%%%%%%%%%%%%%% References %%%%%%%%%%%%%%%%%%%%%%%%%

\bibliography{references}

\begin{thebibliography}{10}
\newcommand{\enquote}[1]{``#1''}

\bibitem{tyson-ao}
R.~Tyson, \emph{Principles of adaptive optics} (CRC press, Boca Raton, 2010).

\bibitem{liu2018observing}
T.-l. Liu, S.~Upadhyayula, D.~E. Milkie, V.~Singh, K.~Wang, I.~A. Swinburne,
  K.~R. Mosaliganti, Z.~M. Collins, T.~W. Hiscock, J.~Shea \emph{et~al.},
  \enquote{Observing the cell in its native state: Imaging subcellular dynamics
  in multicellular organisms,} {\protect\JournalTitle{Science}} \textbf{360},
  eaaq1392 (2018).

\bibitem{dubose2018handheld}
T.~DuBose, D.~Nankivil, F.~LaRocca, G.~Waterman, K.~Hagan, J.~Polans,
  B.~Keller, D.~Tran-Viet, L.~Vajzovic, A.~N. Kuo \emph{et~al.},
  \enquote{Handheld adaptive optics scanning laser ophthalmoscope,}
  {\protect\JournalTitle{Optica}} \textbf{5}, 1027--1036 (2018).

\bibitem{azucena2011adaptive}
O.~Azucena, J.~Crest, S.~Kotadia, W.~Sullivan, X.~Tao, M.~Reinig, D.~Gavel,
  S.~Olivier, and J.~Kubby, \enquote{Adaptive optics wide-field microscopy
  using direct wavefront sensing,} {\protect\JournalTitle{Optics {L}etters}}
  \textbf{36}, 825--827 (2011).

\bibitem{wang2014rapid}
K.~Wang, D.~E. Milkie, A.~Saxena, P.~Engerer, T.~Misgeld, M.~E. Bronner,
  J.~Mumm, and E.~Betzig, \enquote{Rapid adaptive optical recovery of optimal
  resolution over large volumes,} {\protect\JournalTitle{nature methods}}
  \textbf{11}, 625 (2014).

\bibitem{wang2015direct}
K.~Wang, W.~Sun, C.~T. Richie, B.~K. Harvey, E.~Betzig, and N.~Ji,
  \enquote{Direct wavefront sensing for high-resolution in vivo imaging in
  scattering tissue,} {\protect\JournalTitle{Nature communications}}
  \textbf{6}, 7276 (2015).

\bibitem{booth2014adaptive}
M.~J. Booth, \enquote{Adaptive optical microscopy: the ongoing quest for a
  perfect image,} {\protect\JournalTitle{Light: Science \& Applications}}
  \textbf{3}, e165 (2014).

\bibitem{ji2017adaptive}
N.~Ji, \enquote{Adaptive optical fluorescence microscopy,}
  {\protect\JournalTitle{Nature methods}} \textbf{14}, 374 (2017).

\bibitem{kubby-bio-ao}
J.~A. Kubby, \emph{Adaptive optics for biological imaging} (CRC press, Boca
  Raton, 2013).

\bibitem{booth2006wave}
M.~J. Booth, \enquote{Wave front sensor-less adaptive optics: a model-based
  approach using sphere packings,} {\protect\JournalTitle{Optics {E}xpress}}
  \textbf{14}, 1339--1352 (2006).

\bibitem{ji2010adaptive}
N.~Ji, D.~E. Milkie, and E.~Betzig, \enquote{Adaptive optics via pupil
  segmentation for high-resolution imaging in biological tissues,}
  {\protect\JournalTitle{Nature methods}} \textbf{7}, 141 (2010).

\bibitem{linhai2011wavefront}
H.~Linhai and C.~Rao, \enquote{Wavefront sensorless adaptive optics: a general
  model-based approach,} {\protect\JournalTitle{Optics {E}xpress}} \textbf{19},
  371--379 (2011).

\bibitem{vzurauskas2019isosense}
M.~{\v{Z}}urauskas, I.~M. Dobbie, R.~M. Parton, M.~A. Phillips, A.~G{\"o}hler,
  I.~Davis, and M.~J. Booth, \enquote{Isosense: frequency enhanced sensorless
  adaptive optics through structured illumination,}
  {\protect\JournalTitle{Optica}} \textbf{6}, 370--379 (2019).

\bibitem{bagwell2006liquid}
B.~E. Bagwell, D.~V. Wick, R.~Batchko, J.~D. Mansell, T.~Martinez, S.~R.
  Restaino, D.~M. Payne, J.~Harriman, S.~Serati, G.~Sharp \emph{et~al.},
  \enquote{Liquid crystal based active optics,} in \emph{Novel optical systems
  design and optimization IX,}  vol. 6289 (International Society for Optics and
  Photonics, 2006), p. 628908.

\bibitem{tanabe2016transmissive}
A.~Tanabe, T.~Hibi, S.~Ipponjima, K.~Matsumoto, M.~Yokoyama, M.~Kurihara,
  N.~Hashimoto, and T.~Nemoto, \enquote{Transmissive liquid-crystal device for
  correcting primary coma aberration and astigmatism in biospecimen in
  two-photon excitation laser scanning microscopy,}
  {\protect\JournalTitle{Journal of biomedical optics}} \textbf{21}, 121503
  (2016).

\bibitem{bonora2015wavefront}
S.~Bonora, Y.~Jian, P.~Zhang, A.~Zam, E.~N. Pugh, R.~J. Zawadzki, and M.~V.
  Sarunic, \enquote{Wavefront correction and high-resolution in vivo oct
  imaging with an objective integrated multi-actuator adaptive lens,}
  {\protect\JournalTitle{Optics {E}xpress}} \textbf{23}, 21931--21941 (2015).

\bibitem{two-photon-scanning}
J.~M. Bueno, M.~Skorsetz, S.~Bonora, and P.~Artal, \enquote{Wavefront
  correction in two-photon microscopy with a multi-actuator adaptive lens,}
  {\protect\JournalTitle{Optics {E}xpress}} \textbf{26}, 14278--14287 (2018).

\bibitem{philipp2019diffraction}
K.~Philipp, F.~Lemke, S.~Scholz, U.~Wallrabe, M.~C. Wapler, N.~Koukourakis, and
  J.~W. Czarske, \enquote{Diffraction-limited axial scanning in thick
  biological tissue with an aberration-correcting adaptive lens,}
  {\protect\JournalTitle{Scientific reports}} \textbf{9}, 9532 (2019).

\bibitem{dorn2019compact}
A.~Dorn, P.~Rajaeipour, K.~Banerjee, H.~Zappe, and {\c{C}}.~Ataman,
  \enquote{Compact all-in-line adaptive optics fluorescence microscope using an
  optofluidic phase modulator,} in \emph{2019 International Conference on
  Optical MEMS and Nanophotonics (OMN),}  (IEEE, 2019), pp. 178--179.

\bibitem{OFAO-AO}
K.~Banerjee, P.~Rajaeipour, {\c{C}}.~Ataman, and H.~Zappe, \enquote{Optofluidic
  adaptive optics,} {\protect\JournalTitle{Applied {O}ptics}} \textbf{57},
  6338--6344 (2018).

\bibitem{OFAO-control-ao}
P.~Rajaeipour, K.~Banerjee, H.~Zappe, and {\c{C}}.~Ataman,
  \enquote{Optimization-based real-time open-loop control of an optofluidic
  refractive phase modulator,} {\protect\JournalTitle{Applied {O}ptics}}
  \textbf{58}, 1064--1072 (2019).

\bibitem{plossl-config}
A.~Negrean and H.~D. Mansvelder, \enquote{Optimal lens design and use in
  laser-scanning microscopy,} {\protect\JournalTitle{Biomedical {O}ptics
  {E}xpress}} \textbf{5}, 1588--1609 (2014).

\bibitem{OFAO-OMN2018}
K.~Banerjee, P.~Rajaeipour, {\c{C}}.~Ataman, and H.~Zappe, \enquote{An
  optofiuidic refractive phase modulator with an electrostatic 2d actuator
  array,} in \emph{2018 International Conference on Optical MEMS and
  Nanophotonics (OMN),}  (IEEE, 2018), pp. 1--2.

\bibitem{thibos2002standards}
L.~N. Thibos, R.~A. Applegate, J.~T. Schwiegerling, and R.~Webb,
  \enquote{Standards for reporting the optical aberrations of eyes,}
  {\protect\JournalTitle{Journal of refractive surgery}} \textbf{18},
  S652--S660 (2002).

\bibitem{booth-lsf}
D.~D{\'e}barre, M.~J. Booth, and T.~Wilson, \enquote{Image based adaptive
  optics through optimisation of low spatial frequencies,}
  {\protect\JournalTitle{Optics {E}xpress}} \textbf{15}, 8176--8190 (2007).

\bibitem{booth2002adaptive}
M.~J. Booth, M.~A. Neil, R.~Ju{\v{s}}kaitis, and T.~Wilson, \enquote{Adaptive
  aberration correction in a confocal microscope,}
  {\protect\JournalTitle{Proceedings of the National Academy of Sciences}}
  \textbf{99}, 5788--5792 (2002).

\bibitem{sensorless-accuracy}
A.~Facomprez, E.~Beaurepaire, and D.~D{\'e}barre, \enquote{Accuracy of
  correction in modal sensorless adaptive optics,}
  {\protect\JournalTitle{Optics {E}xpress}} \textbf{20}, 2598--2612 (2012).

\bibitem{kazasidis2018algorithm}
O.~Kazasidis, S.~Verpoort, and U.~Wittrock, \enquote{Algorithm design for
  image-based wavefront control without wavefront sensing,} in \emph{Optical
  Instrument Science, Technology, and Applications,}  vol. 10695 (International
  Society for Optics and Photonics, 2018), p. 1069502.

\bibitem{mtf-measurement}
G.~D. Boreman, \emph{Modulation transfer function in optical and
  electro-optical systems}, vol.~21 (SPIE press Bellingham, WA, 2001).

\bibitem{OFAO-PW-pushpull}
K.~Banerjee, P.~Rajaeipour, H.~Zappe, and {\c{C}}.~Ataman, \enquote{Refractive
  opto-fluidic wavefront modulator with electrostatic push-pull actuation,} in
  \emph{Adaptive Optics and Wavefront Control for Biological Systems V,}  vol.
  10886 (International Society for Optics and Photonics, 2019), p. 108860D.

\bibitem{schwertner2004characterizing}
M.~Schwertner, M.~J. Booth, and T.~Wilson, \enquote{Characterizing specimen
  induced aberrations for high {NA} adaptive optical microscopy,}
  {\protect\JournalTitle{Optics express}} \textbf{12}, 6540--6552 (2004).

\end{thebibliography}

\end{document}